\documentclass[aps,prb,twocolumn,showpacs,superscriptaddress]{revtex4}

\usepackage{graphicx}
\usepackage{amsmath}
\usepackage{amssymb}

\begin{document}

\title{Momentum and doping dependence of spin excitations in electron-doped cuprate superconductors}

\author{Pengfei Jing}

\affiliation{Department of Physics, Beijing Normal University, Beijing 100875, China}

\author{Huaisong Zhao}

\affiliation{College of Physics, Qingdao University, Qingdao 266071, China}

\author{L\"ulin Kuang}

\affiliation{Sugon National Research Center for High-performance Computing Engineering Technology, Beijng 100093, China}

\author{Yu Lan}

\affiliation{College of Physics and Electronic Engineering, Hengyang Normal University, Hengyang 421002, China}

\author{Shiping Feng}
\email{spfeng@bnu.edu.cn}

\affiliation{Department of Physics, Beijing Normal University, Beijing 100875, China}

\begin{abstract}
Superconductivity in copper oxides emerges on doping holes or electrons into their Mott insulating parent compounds. The spin excitations are thought to be the mediating glue for the pairing in superconductivity. Here the momentum and doping dependence of the dynamical spin response in the electron-doped cuprate superconductors is studied based on the kinetic-energy-driven superconducting mechanism. It is shown that the dispersion of the low-energy spin excitations changes strongly upon electron doping, however, the hour-glass-shaped dispersion of the low-energy spin excitations appeared in the hole-doped side is absent in the electron-doped case due to the electron-hole asymmetry. In particular, the commensurate resonance appears in the superconducting-state with the resonance energy that correlates with the dome-shaped doping dependence of the superconducting gap. Moreover, the spectral weight and dispersion of the high-energy spin excitations in the superconducting-state are comparable with those in the corresponding normal-state, indicating that the high-energy spin excitations do not play an important part in the pair formation.
\end{abstract}


\maketitle

\section{Introduction}\label{Introduction}
Cuprate superconductors are separated into two groups: the hole-doped and electron-doped cuprate superconductors \cite{Bednorz86,Tokura89}, respectively. This follows a fact that the undoped parent compounds of cuprate superconductors are known to be a Mott insulator with an antiferromagnetic (AF) long-range order (AFLRO) \cite{Vaknin87}. However, this AFLRO is destructed quickly by doping holes or electrons, and then superconductivity emerges leaving the AF short-range order (AFSRO) correlation still intact \cite{Kastner98,Armitage10,Fujita12}. Therefore the persistence of spin excitations is apparently universal in both hole- and electron-doped cuprate superconductors \cite{Kastner98,Armitage10,Fujita12}. Immediately following the discovery of superconductivity in cuprate superconductors, it has been realized that the spin excitations may serve as the pairing glue \cite{Anderson87}. In this case, the understanding what survives of the spin excitations in the doped regime is crucially important for the understanding of the emergence of superconductivity.

The early inelastic neutron scattering (INS) measurements have demonstrated that the spin excitation of the undoped parent compounds is well described by spin-wave theory \cite{Vaknin87}. However, when holes are doped, the low-energy spin excitation changes substantially and has an hour-glass-shaped dispersion in the superconducting (SC) state \cite{Birgeneau89,Arai99,Pailhes06,Xu09}, where two incommensurate (IC) components of the low-energy spin excitation spectrum are separated by a commensurate resonance energy $\omega_{\rm r}$ at the waist of the hour glass. In particular, in analogy to the dome-shaped doping dependence of the SC transition temperature $T_{\rm c}$, this commensurate resonance energy $\omega_{\rm r}$ increases with increasing doping in the underdoped regime, and reaches a maximum around the optimal doping, then decreases in the overdoped regime. However, very recently, the resonant inelastic X-ray scattering (RIXS) experimental observations \cite{Dean14,Tacon11,Dean13} indicate that although the spectral distribution of the high-energy spin excitations broadens upon hole doping, it keeps its energy position almost unchanged, i.e., it follows rather closely the spin excitation dispersion of the parent compound, reflecting a fact that the high-energy spin excitation is a remnant mode of the parent compound and then the spin dynamics has strongly localized nature.

On the electron-doped side, although the low-energy spin excitation spectrum in the upward component, above $\omega_{\rm r}$, is the similar that in the hole-doped case, the low-energy IC magnetic scattering below $\omega_{\rm r}$ and inward dispersion toward a resonance peak with increasing energy appeared in the hole-doped case are not observed  \cite{Yamada03,Wilson06,Fujita06,Yu09,Zhao11}. However, as in the hole-doped case \cite{Pailhes06}, $\omega_{\rm r}$ in the electron-doped side still scales with the SC gap forming a universal plot \cite{Yu09}. Moreover, the recent RIXS experimental data show that the spin excitations of the parent compound shift to higher energy upon electron doping over a wide momentum space, and the spin excitations have the unusually large hardening of the dispersion, indicating that the spin excitations in the electron-doped side more deeply couple to the charge carriers \cite{Ishii14,Lee14}.

Although the similarities and differences of the spin excitation spectrum between the electron- and hole-doped cuprate superconductors from low-energy to high-energy have been observed from the INS and RIXS experiments \cite{Birgeneau89,Arai99,Pailhes06,Xu09,Dean14,Tacon11,Dean13,Yamada03,Wilson06,Fujita06,Yu09,Zhao11,Ishii14,Lee14}, the full understanding of their similarities and differences is still a challenging issue. Theoretically, there is a general consensus that the unusual dynamical spin response is generated by the scattering of spins due to the charge-carrier fluctuations. Several attempts have been made to make these arguments more precise \cite{Li16,Kuang15,Jia14,Brinckmann99,Ismer07}. In particular, it has been shown that the observed low-energy resonance peak in the electron-doped cuprate superconductors is likely an overdamped spin exciton located near the particle-hole continuum \cite{Ismer07}. The dynamical spin response is manifested itself by the dynamical spin structure factor $S({\bf k},\omega)$. In our early discussions of the origin of the low-energy commensurate resonance in the electron-doped cuprate superconductors, $S({\bf k},\omega)$ was calculated in terms of the collective charge-carrier mode in the particle-particle channel only, and then the low-energy commensurate resonance has been identified as a nature consequence of the creation of the charge-carrier pairs \cite{Cheng08}. However, an obvious weakness is that the contribution of the collective charge-carrier mode in the particle-hole channel to the spin dynamics is not considered, and therefore the formalism in Ref. \cite{Cheng08} can not be used directly to discuss the dynamical spin response in the normal-state. Moreover, although this contribution of the collective charge-carrier mode in the particle-hole channel to the spin dynamics is not related with the low-energy commensurate resonance, it contributes to the low-energy IC spin fluctuations and high-energy spin excitation spectrum. In this paper, we study the dynamical spin response of the electron-doped cuprate superconductors in both the SC- and normal-states by including the contributions of the collective modes in the particle-particle and particle-hole channels. As a complement of the our previous analysis of the nature of the low-energy commensurate resonance in the SC-state \cite{Cheng08}, we first discuss the low-energy behavior of the dynamical spin response, and then confirm that the low-energy commensurate resonance energy correlates with the dome-shaped doping dependence of the charge-carrier pair gap. However, the high-energy spin excitations bear a striking resemblance to those obtained in the corresponding normal-state, indicating that the high-energy spin excitations are unlikely to be a major factor in the pairing interaction.

The rest of this paper is organized as follows. The basic formalism is presented in Sec. \ref{Formalism}, where we generalize the calculation of the dynamical spin structure factor  in Ref. \onlinecite{Cheng08} by including the contributions of the collective modes in both the particle-particle and particle-hole channels. Within this theoretical framework, we discuss the momentum and doping dependence of the dynamical spin response for the electron-doped cuprate superconductors in Sec. \ref{sc-state}, where we show that the hour-glass-shaped dispersion of the low-energy spin excitations appeared in the hole-doped case is absent in the electron-doped side. However, the low-energy IC spin fluctuations can persist into the normal-state. Finally, we give a summary and discussions in Sec. \ref{conclusions}.

\section{Formalism}\label{Formalism}

We start from the $t$-$J$ model on a square lattice, which is the simplest model capturing the essential physics of cuprate superconductors \cite{Anderson87}. This $t$-$J$ model describes a competition between the kinetic energy and magnetic energy, where the magnetic energy with the nearest-neighbor (NN) spin-spin antiferromagnetic (AF) exchange $J$ favors the magnetic order for spins and results in frustration of the kinetic energy, while the kinetic energy generally includes the electron NN hopping $t$ and next NN hopping $t'$, and therefore it favors delocalization of charge carriers and tends to destroy the magnetic order. In this paper, the parameters are chosen as $t/J=-2.5$, $t'/t=0.3$. When necessary to compare with the experimental data, we take $J=110$ meV \cite{Vaknin87}. As in the hole-doped case \cite{Yu92,Dagotto94,Lee99}, the electron-doped $t$-$J$ model is also very difficult to analyze, analytically as well as numerically, because of the restriction of the motion of electrons in the restricted Hilbert space without zero electron occupancy. However, this restriction of no zero occupancy can be treated properly within the framework of the fermion-spin theory \cite{Feng9404,Feng15}, where the spin fluctuations occur in the charge-carrier quasiparticle background, and then the spin configuration in the $t$-$J$ model is strongly rearranged due to the effect of the charge-carrier hopping on spins.

For the discussions of the dynamical spin response of the electron-doped cuprate superconductors, we need to calculate the dynamical spin structure factor $S({\bf k}, \omega)$ of the $t$-$J$ model, which is related directly to the spin Green's function as $S({\bf k},\omega)=-2[1+n_{\rm B}(\omega)]{\rm Im}D({\bf k},\omega)$, with the boson distribution function $n_{\rm B}(\omega)$ and the spin Green's function $D({\bf k},\omega)$ that is defined as $D(l-l',t-t')=\langle\langle S^{+}_{l}(t);S^{-}_{l'}(t') \rangle\rangle$. Based on the $t$-$J$ model in the fermion-spin representation \cite{Feng9404,Feng15}, the kinetic-energy-driven superconductivity has been developed \cite{Feng15,Feng0306}, where the interaction between charge carriers and spins directly from the kinetic energy by the exchange of spin excitations induces the SC-state in the particle-particle channel, and then the charge-carrier pair gap parameter and $T_{\rm c}$ have a dome-shaped doping dependence. Within this kinetic-energy-driven SC mechanism, the doping dependence of the low-energy commensurate resonance in the electron-doped cuprate superconductors in the SC-state has been discussed by considering the contribution of the collective charge-carrier mode in the particle-particle channel \cite{Cheng08}. Following these previous discussions \cite{Cheng08}, the SC-state dynamical spin structure factor of the electron-doped $t$-$J$ model in the fermion-spin representation can be obtained as,
\begin{eqnarray}\label{DSSF}
S({\bf k},\omega)&=&{-2[1+n_{\rm B}(\omega)]B^{2}_{{\bf k}}{\rm Im}\Sigma^{({\rm s})}({\bf k},\omega)\over [\omega^{2}-\omega^{2}_{\bf k}-B_{{\bf k}} {\rm Re}\Sigma^{({\rm s})}({\bf k}, \omega)]^{2}+[B_{{\bf k}}{\rm Im}\Sigma^{({\rm s})}({\bf k},\omega)]^{2}}, \nonumber\\
&~&~
\end{eqnarray}
with ${\rm Im}\Sigma^{({\rm s})}({\bf k},\omega)$ and ${\rm Re}\Sigma^{({\rm s})}({\bf k},\omega)$ that are the corresponding imaginary and real parts of the spin self-energy $\Sigma^{({\rm s})}({\bf k},\omega)$, respectively, while the spin self-energy $\Sigma^{({\rm s})}({\bf k},\omega)$ can be separated into two parts as:
\begin{eqnarray}\label{SSF}
\Sigma^{({\rm s})}({\bf k},\omega)&=&\Sigma^{({\rm s})}_{\rm ph}({\bf k},\omega)+\Sigma^{({\rm s})}_{\rm pp}({\bf k},\omega),
\end{eqnarray}
where $\Sigma^{({\rm s})}_{\rm ph}({\bf k},\omega)$ comes from the spin fluctuations in the mobile charge-carrier quasiparticle background, and can be evaluated explicitly in terms of the collective mode in the particle-hole channel as,
\begin{eqnarray}\label{SSF-ph}
\Sigma^{({\rm s})}_{\rm ph}({\bf k},\omega)&=&{1\over 2N^2}\sum_{{\bf pq},\nu=1,2}(-1)^{\nu}\Omega_{\rm a}({\bf k},{\bf p},{\bf q})\nonumber\\ &\times& {I^{({\rm a})}_{+} ({\bf p},{\bf q})F_{\nu+}^{({\rm s})}({\bf k},{\bf p},{\bf q})\over\omega^{2}-[\omega_{{\bf q}+{\bf k}}-(-1)^{\nu+1} (E_{{\rm a}{\bf p}+{\bf q}}-E_{{\rm a}{\bf p}})]^{2}},~~~~
\end{eqnarray}
with $\Omega_{\rm a}({\bf k},{\bf p},{\bf q})=Z^{2}_{\rm aF}(\Lambda^{2}_{{\bf k}-{\bf p}}+\Lambda^{2}_{{\bf p}+{\bf q}+{\bf k}})B_{{\bf q}+{\bf k}}/(2\omega_{{\bf q}+{\bf k}})$,
$\Lambda_{{\bf k}}=4t\gamma_{\bf k}-4t'\gamma_{\bf k}'$, $\gamma_{{\bf k}}=(\cos k_{x}+\cos k_{y})/2$, $\gamma_{{\bf k}}'=\cos k_{x}\cos k_{y}$, the related charge-carrier coherence factors for this process,
\begin{eqnarray}
I^{({\rm a})}_{+}({\bf p},{\bf q})&=& 1+{\bar{\xi}_{\bf p}\bar{\xi}_{{\bf p}+{\bf q}}-\bar{\Delta}_{\rm aZ}({\bf p})\bar{\Delta}_{\rm aZ}({{\bf p}+{\bf q}})\over E_{{\rm a}{\bf p}} E_{{\rm a} {\bf p}+{\bf q}}},~~~~~\label{CFS1}
\end{eqnarray}
and the function,
\begin{eqnarray}
&~&F_{\nu+}^{({\rm s})}({\bf k},{\bf p},{\bf q})=[\omega_{{\bf q}+{\bf k}}-(-1)^{\nu+1}(E_{{\rm a}{\bf p}+{\bf q}}-E_{{\rm a}{\bf p}})]\nonumber\\ &\times& \{n_{\rm B}(\omega_{{\bf q}+{\bf k}})[n_{\rm F} (E_{{\rm a}{\bf p}})-n_{\rm F}(E_{{\rm a}{\bf p}+{\bf q}})]\nonumber\\
&-&(-1)^{\nu+1}n_{\rm F}[(-1)^{\nu}E_{{\rm a}{\bf p}}]n_{\rm F}[(-1)^{\nu+1}E_{{\rm a}{\bf p}+{\bf q}}]\},~~~~~
\end{eqnarray}
while $\Sigma^{({\rm s})}_{\rm pp}({\bf k},\omega)$ comes from the spin fluctuations in the charge-carrier pair background, and is obtained in terms of the collective mode in the particle-particle channel as \cite{Cheng08},
\begin{eqnarray}\label{SSF-pp}
\Sigma^{({\rm s})}_{\rm pp}({\bf k},\omega)&=&{1\over 2N^2}\sum_{{\bf pq},\nu=1,2}(-1)^{\nu}\Omega_{\rm a}({\bf k},{\bf p},{\bf q})\nonumber\\ &\times& {I^{({\rm a})}_{-}({\bf p},{\bf q})F_{\nu-}^{({\rm s})}({\bf k},{\bf p},{\bf q})\over \omega^{2}-[\omega_{{\bf q}+{\bf k}}-(-1)^{\nu+1} (E_{{\rm a}{{\bf p}+{\bf q}}}+ E_{{\rm a}{\bf p}})]^{2}},~~~~
\end{eqnarray}
with the related charge-carrier coherence factors for this process of the creation of charge-carrier pairs,
\begin{eqnarray}
I^{({\rm a})}_{-}({\bf p},{\bf q})&=& 1-{\bar{\xi}_{\bf p}\bar{\xi}_{{\bf p}+{\bf q}}-\bar{\Delta}_{\rm aZ}({\bf p})\bar{\Delta}_{\rm aZ}({{\bf p}+{\bf q}})\over E_{{\rm a}{\bf p}} E_{{\rm a}{\bf p}+{\bf q}}},~~~~~\label{CFS2}
\end{eqnarray}
and the function,
\begin{eqnarray}
&~&F_{\nu-}^{({\rm s})}({\bf k},{\bf p},{\bf q})=[\omega_{{\bf q}+{\bf k}}-(-1)^{\nu+1}(E_{{\rm a}{\bf p}+{\bf q}}+E_{{\rm a}{\bf p}})]\nonumber\\ &\times&\{n_{\rm B}(\omega_{{\bf q}+{\bf k}})[1-n_{\rm F} (E_{{\rm a}{\bf p}})-n_{\rm F}(E_{{\rm a}{\bf p}+{\bf q}})]\nonumber\\
&-&(-1)^{\nu+1}n_{\rm F}[(-1)^{\nu+1}E_{{\rm a}{\bf p}}]n_{\rm F}[(-1)^{\nu+1}E_{{\rm a}{\bf p}+{\bf q}}]\}, ~~~~
\end{eqnarray}
where $n_{\rm F}(\omega)$ is the fermion distribution functions, the charge-carrier quasiparticle spectrum $E_{{\rm a}{\bf k}}=\sqrt{\bar{\xi}^{2}_{{\bf k}}+\mid\bar{\Delta}_{\rm aZ} ({\bf k})\mid^{2}}$, $\bar{\Delta}_{\rm aZ}({\bf k})=Z_{\rm aF}\bar{\Delta}_{\rm a}({\bf k})$, and the charge-carrier pair gap $\bar{\Delta}_{\rm a}({\bf k} )$ has a nonmonotonic d-wave form \cite{Cheng08}, i.e.,
\begin{eqnarray}\label{gap}
\bar{\Delta}_{\rm a}({\bf k})=\bar{\Delta}_{\rm a}[\gamma^{(d)}_{\bf k}+B\gamma^{(2d)}_{\bf k}],
\end{eqnarray}
with $\gamma^{(d)}_{\bf k} =[{\rm cos}k_{x}-{\rm cos} k_{y}]/2$, $\gamma^{(2d)}_{\bf k}=[{\rm cos}(2k_{x})-{\rm cos}(2k_{y})]/2$, and then the maximum charge-carrier pair gap appears not at the Brillouin zone (BZ) boundary as expected from the monotonic d-wave gap, but at the hot spot between [$\pi$,0] and [$\pi/2$,$\pi/2$], in agreement with the angle-resolved photoemission spectroscopy experimental results \cite{Matsui05}, while the mean-field (MF) charge-carrier excitation spectrum $\bar{\xi}_{{\bf k}}$, the charge-carrier quasiparticle coherent weight $Z_{\rm aF}$, the gap parameters $\bar{\Delta}_{\rm a}$ and $B$, the MF spin excitation spectrum $\omega_{\bf k}$, and the function $B_{\bf k}$ have been given in Ref. \onlinecite{Cheng08}. In particular, the gap parameters $\bar{\Delta}_{\rm a}$, $B$, the charge-carrier quasiparticle coherent weight $Z_{\rm aF}$, the chemical potential $\mu$, and other order parameters in the above calculation have been determined by self-consistent calculation \cite{Cheng08} without using any adjustable parameters.

The poles of the dynamical spin structure factor $S({\bf k},\omega)$ in Eq. (\ref{DSSF}) map the energy versus momentum dependence of the spin excitations, i.e., the spin excitation energies are obtained by the solution of the self-consistent equation,
\begin{eqnarray}\label{SEE}
E^{2}_{\rm s}({\bf k})=\omega^{2}_{\bf k}+B_{\bf k}{\rm Re}\Sigma^{({\rm s})}[{\bf k},E_{\rm s}({\bf k})],
\end{eqnarray}
and then these energies can be measured in INS and RIXS experiments \cite{Yamada03,Wilson06,Fujita06,Yu09,Zhao11,Ishii14,Lee14}. On the other hand, $S({\bf k},\omega)$ shows peaks when the incoming scattering energy $\omega$ is equal to the renormalized spin excitation energy $E_{\rm s}({\bf k})$, i.e., $\omega^{2}-E^{2}_{\rm s}({\bf k}_{\rm c})\sim 0$, for the critical wave vectors ${\bf k}_{\bf c}$, the magnetic scattering peaks appear, while the height of these peaks are dominated by the damping (the inverse of the imaginary part of the spin self-energy $1/{\rm Im}\Sigma^{({\rm s})}[{\bf k}_{\rm c},E_{\rm s}({\bf k}_{\rm c})]$).

\section{Evolution of spin excitations with momentum and doping}\label{sc-state}

\begin{figure}[h!]
\centering
\includegraphics[scale=0.33]{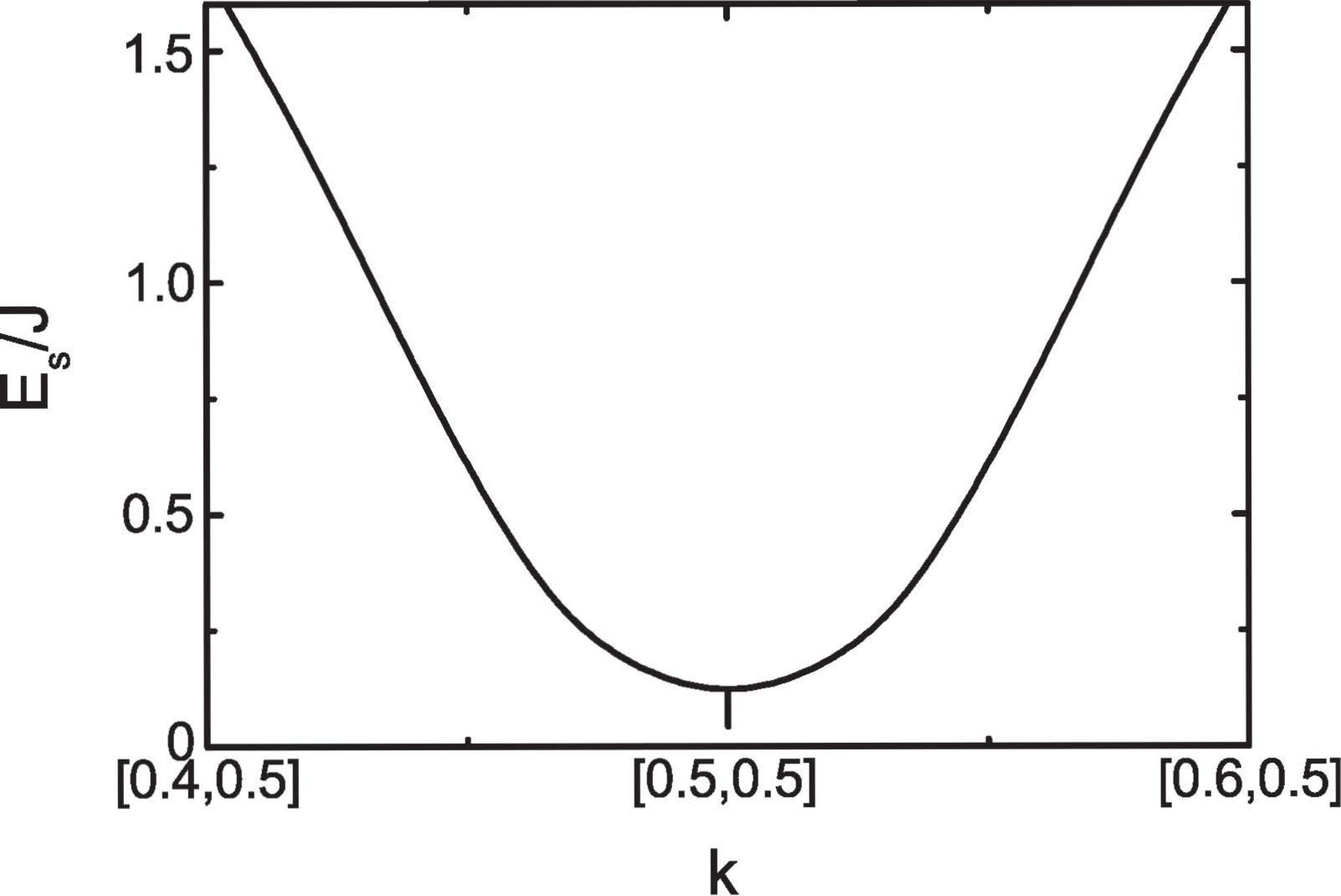}
\caption{The momentum dependence of the position of the low-energy magnetic scattering peaks at $\delta=0.15$ with $T=0.002J$ for $t/J=-2.5$ and $t'/t=0.3$. \label{position}}
\end{figure}

\begin{figure}[h!]
\centering
\includegraphics[scale=0.33]{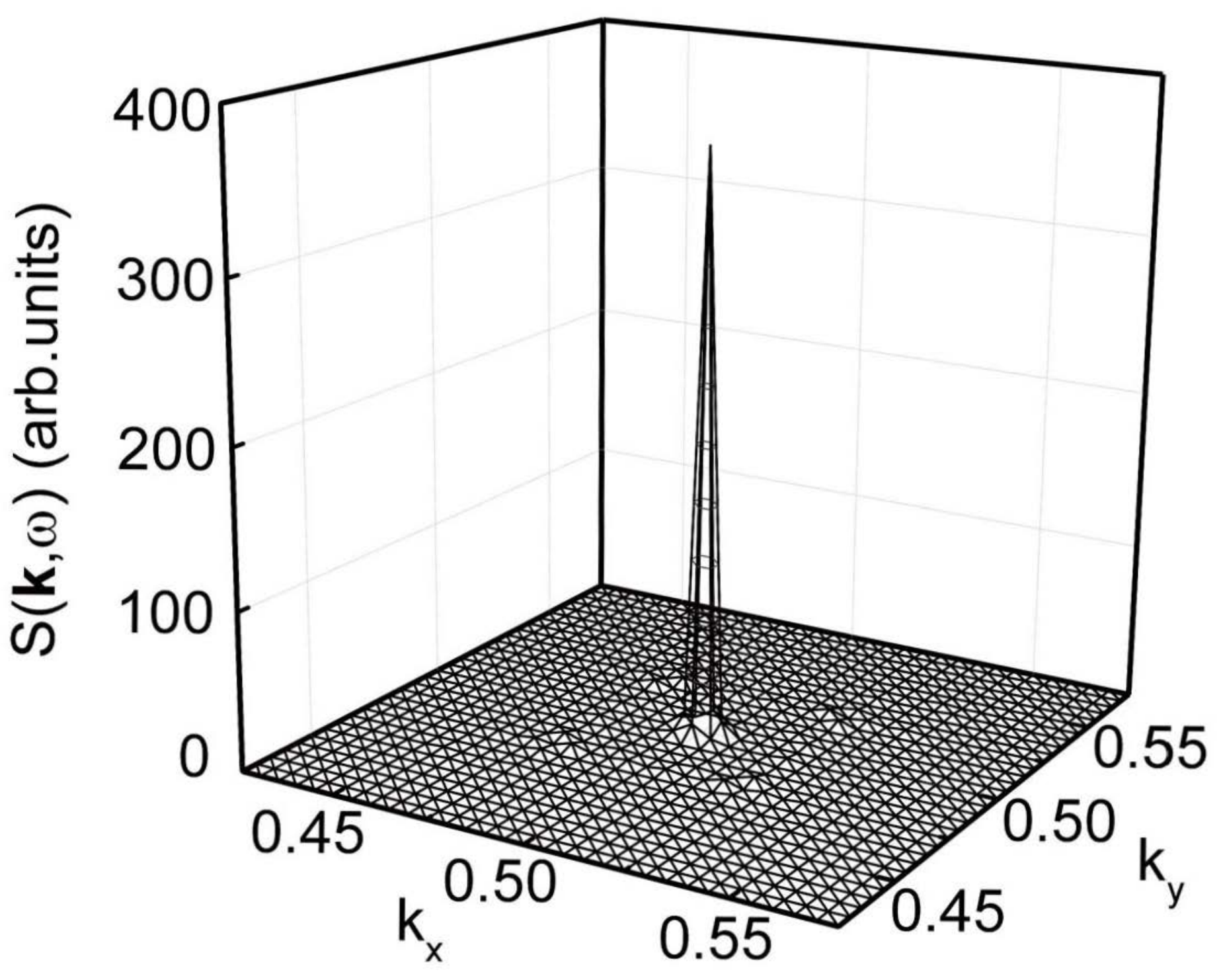}
\caption{The dynamical spin structure factor $S({\bf k},\omega)$ in the $[k_{x},k_{y}]$ plane at $\delta=0.15$ with $T=0.002J$ for $t/J=-2.5$ and $t'/t=0.3$ in $\omega=0.094J$. \label{resonance-energy}}
\end{figure}

\begin{figure*}[t!]
\centering
\includegraphics[scale=0.5]{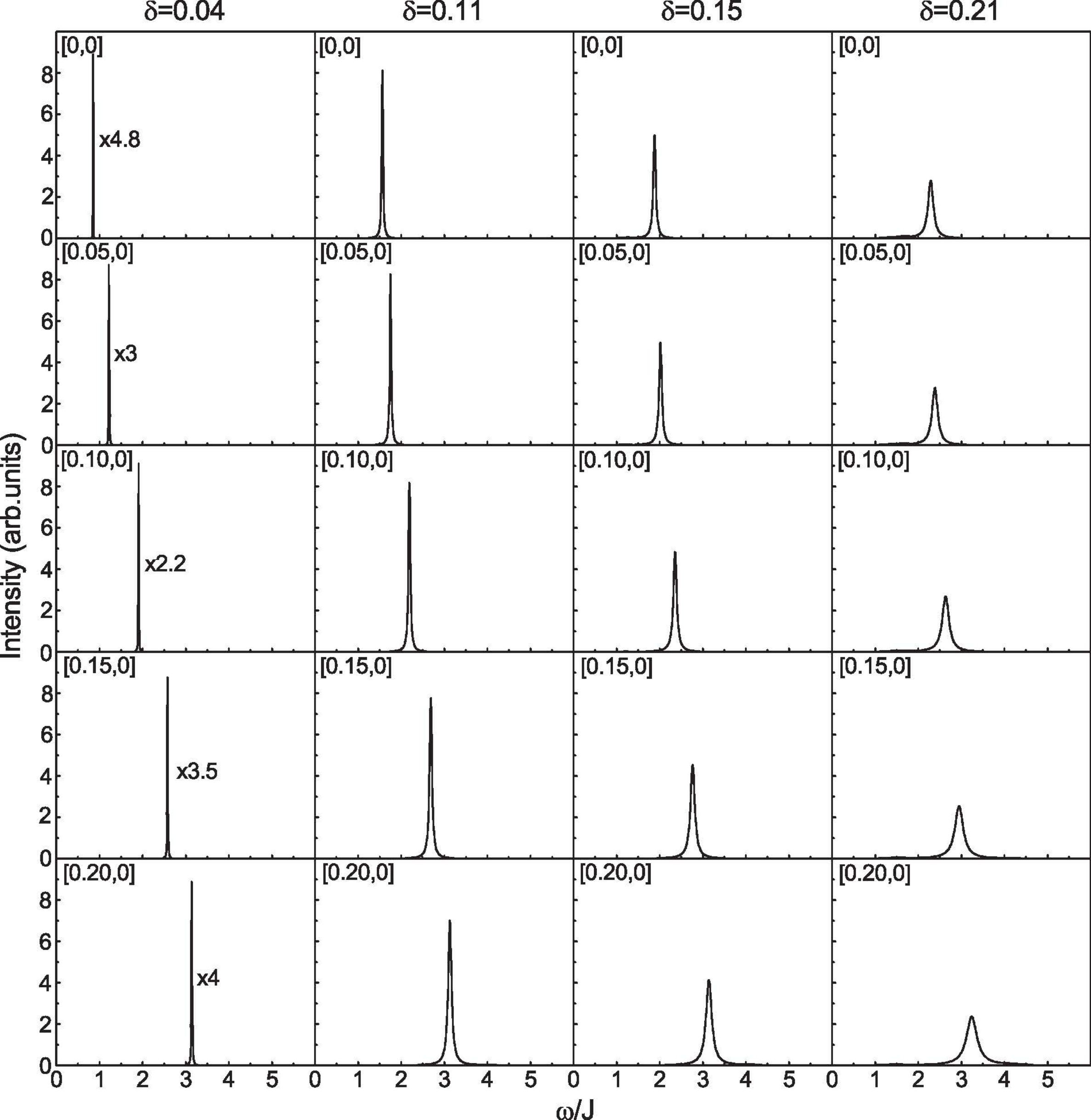}
\caption{The dynamical spin structure factor $S({\bf k},\omega)$ as a function of energy along the ${\bf k}=[0,0]$ to ${\bf k}=[0.5,0]$ direction of the Brillouin zone at $\delta=0.04$, $\delta=0.11$, $\delta=0.15$, and $\delta=0.21$ with $T=0.002J$ for $t/J=-2.5$ and $t'/t=0.3$. \label{SC-high-energy}}
\end{figure*}

We are now ready to discuss the momentum and doping dependence of the spin excitations in the electron-doped cuprate superconductors. Firstly, we study the low-energy dynamical spin response. At half-filling, the $t$-$J$ model is reduced to an AF Heisenberg model, and then the dynamical spin structure factor in Eq. (\ref{DSSF}) is reduced as,
\begin{eqnarray}\label{DSSF0}
S({\bf k},\omega)=\pi {B^{(0)}_{\bf k}\over\omega^{(0)}_{\bf k}}[1+n_{\rm B}(\omega)][\delta(\omega-\omega^{(0)}_{\bf k})- \delta(\omega+\omega^{(0)}_{\bf k})], ~~~
\end{eqnarray}
with the spin-wave spectrum $\omega^{(0)}_{\bf k}=4.75J\sqrt{1-\gamma^{2}_{\bf k}}$ and the function $B^{(0)}_{\bf k}=3.31J(1-\gamma_{\bf k})$. In this case, the position of the lowest energy magnetic scattering peak locates at the AF wave vector $[1/2,1/2]$ (hereafter we use the units of $[2\pi,2\pi]$), so the commensurate AF peak appears there. However, the spin excitation spectrum is strongly renormalized by electron doping, and then the position of the commensurate magnetic scattering peak moves towards to higher energy. To show this point clearly, we have performed a calculation for the self-consistent equation (\ref{SEE}) around $[1/2,1/2]$ of BZ, and the result of the positions of the low-energy magnetic scattering peaks as a function of momentum with temperature $T=0.002J$ at electron doping $\delta=0.15$ is plotted in Fig. \ref{position}, where the self-consistently obtained value of the charge-carrier pair gap parameter is $\bar{\Delta}_{a}=0.051J$. It is shown clearly that there is a broad commensurate magnetic scattering at energies $\omega\leq 0.094J$. For a better understanding of the commensurate magnetic scattering, we plot $S({\bf k},\omega)$ in the ($k_{x},k_{y}$) plane at $\delta=0.15$ for $\omega=0.094J$ with $T=0.002J$ in Fig. \ref{resonance-energy}, where a sharp commensurate magnetic scattering peak emerges at the resonance energy $\omega_{\rm r}=0.094J=10.3$ meV, which is well consistent with the resonance energy $\approx 10.5$ meV observed in Pr$_{0.88}$LaCe$_{0.12}$CuO$_{4-\delta}$ \cite{Wilson06,Yu09}. In particular, we find that $\omega_{\rm r}\approx 2\bar{\Delta}_{a}$, reflecting a fact that the commensurate resonance is universally related to the charge-carrier pair gap, in good agreement with the experimental results \cite{Yu09}. Furthermore, as in the previous discussions \cite{Cheng08}, we have also made a series of calculations for the resonance energy at different doping, and the result shows that in analogy to the doping dependence of the charge-carrier pair gap parameter and $T_{\rm c}$ \cite{Peng97}, $\omega_{\rm r}$ has a dome-shaped doping dependence. On the other hand, the result in Fig. \ref{position} also indicates that above the broad commensurate magnetic scattering, the IC magnetic scattering appears, which is the same as that in the hole-doped case. In particular, the low-energy spin excitations above $\omega_{\rm r}$ have a dispersion similar to the spin-wave, in qualitative agreement with the INS experimental data \cite{Yamada03,Wilson06,Fujita06,Yu09,Zhao11}. However, below $\omega_{\rm r}$, the IC magnetic scattering is absence, which is different from the hole-doped case, where the IC magnetic scattering peaks with the anisotropic distribution of the spectral weight are obtained \cite{Kuang15}, indicating that impact of doping on the dispersion of the low-energy spin excitations around $[1/2,1/2]$ is not symmetric. In other words, in contrast to the hour-glass shaped dispersion observed in the hole-doped case \cite{Arai99,Pailhes06}, the dispersion of the spin excitations in the electron-doped side is a similar spin-wave response centered around the commensurate resonance position \cite{Yamada03,Wilson06,Fujita06,Yu09,Zhao11}. Furthermore, the present result in Fig. \ref{position} also shows that a doping dependence of the spin gap exists in the spin excitation spectrum. In particular, the obtained value of the spin gap at the underdoping $\delta=0.11$ is $0.041J=4.5$ meV, closely matching the experimental value of the spin gap 5 meV found in Nd$_{1.85}$Ce$_{0.15}$CuO$_{4-\delta}$ in the underdoped regime \cite{Ismer07}.

\begin{figure}[h!]
\centering
\includegraphics[scale=0.33]{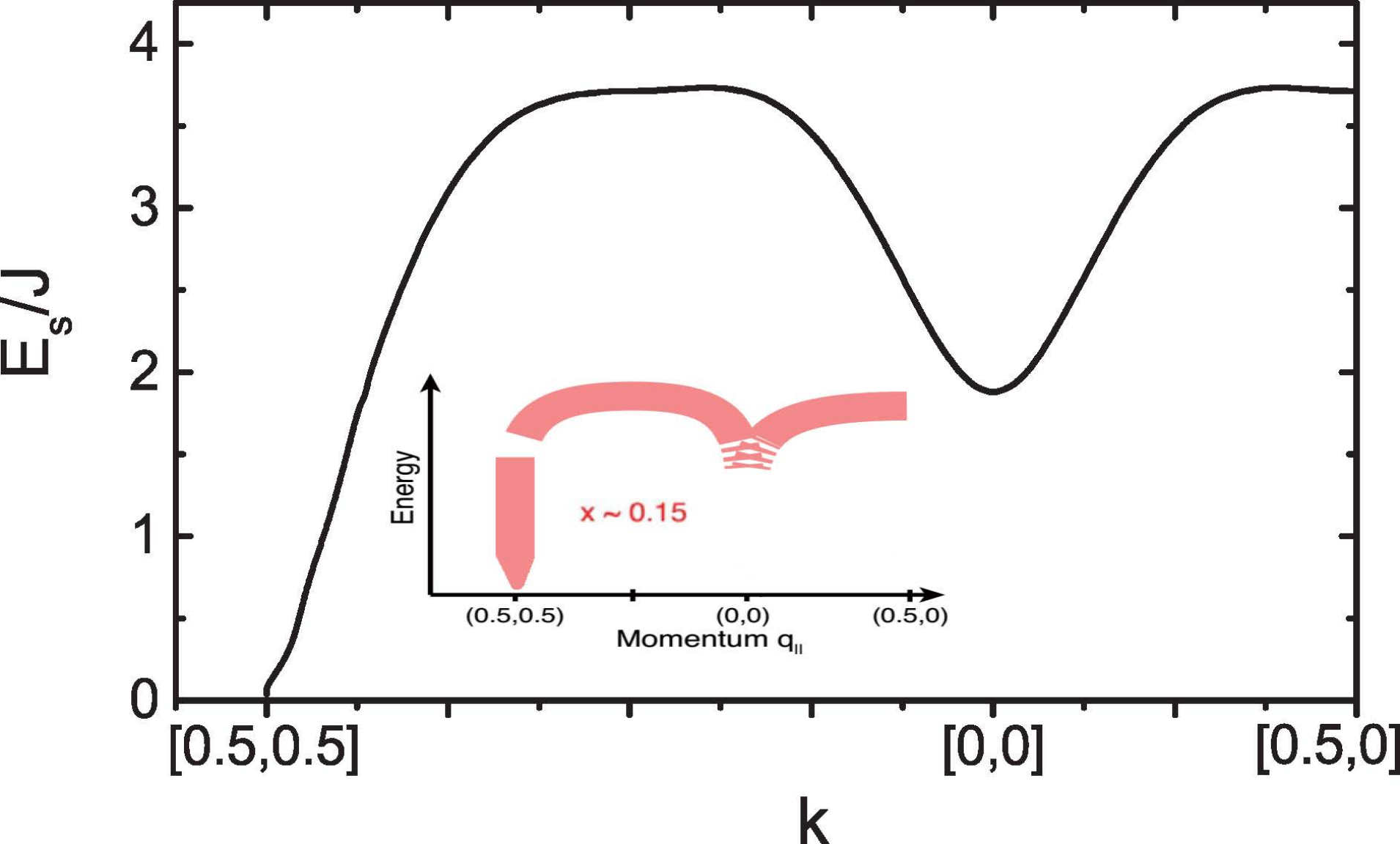}
\caption{(Color online) The dispersion of the spin excitations along the high symmetry directions of the Brillouin zone at $\delta=0.15$ with $T=0.002J$ for $t/J=-2.5$ and $t'/J=0.3$. Inset: the schematic summary of the spin excitation dispersion of the electron-doped cuprate superconductors obtained by resonant inelastic X-ray scattering and inelastic neutron scattering experiments taken from Ref. \onlinecite{Ishii14}. \label{dispersion}}
\end{figure}

Now we turn to discuss the remarkable features of the high-energy magnetic scattering. In Fig. \ref{SC-high-energy}, we plot $S({\bf k},\omega)$ as a function of energy along the ${\bf k}=[0,0]$ to ${\bf k}=[0.5,0]$ direction of BZ at $\delta=0.04$, $\delta=0.11$, $\delta=0.15$, and $\delta=0.21$ with $T=0.002J$. Apparently, the main feature of the high-energy spin excitations in the electron-doped cuprate superconductors \cite{Ishii14} is qualitatively reproduced, where the peak width of the spin excitation broaden upon electron doping. In particular, the spectral weight around $[0,0]$ moves to higher energy with the increase of electron doping. This high-energy shift and broadness of the spin excitations around $[0,0]$ upon electron doping are consistent with the experimental observations \cite{Ishii14}. In Fig. \ref{dispersion}, we plot the spin excitation dispersion along the high symmetry directions of BZ at $\delta=0.15$ with $T=0.002J$ to summarize our main results of the dynamical spin response of the electron-doped cuprate superconductors in the SC-state from low-energy to high-energy. For comparison, the schematic summary of the spin excitation dispersion of the electron-doped cuprate superconductors obtained by RIXS and INS measurements \cite{Ishii14} is also shown in Fig. \ref{dispersion} (inset). It is thus shown that the spin excitations in the electron-doped cuprate superconductors are well defined at all momenta, however, as in the hole-doped case, the low-energy spin excitations depend sensitively on electron doping and momentum, while the electron doping has a more modest effect on the high-energy spin excitations.

\begin{figure}[h!]
\centering
\includegraphics[scale=0.33]{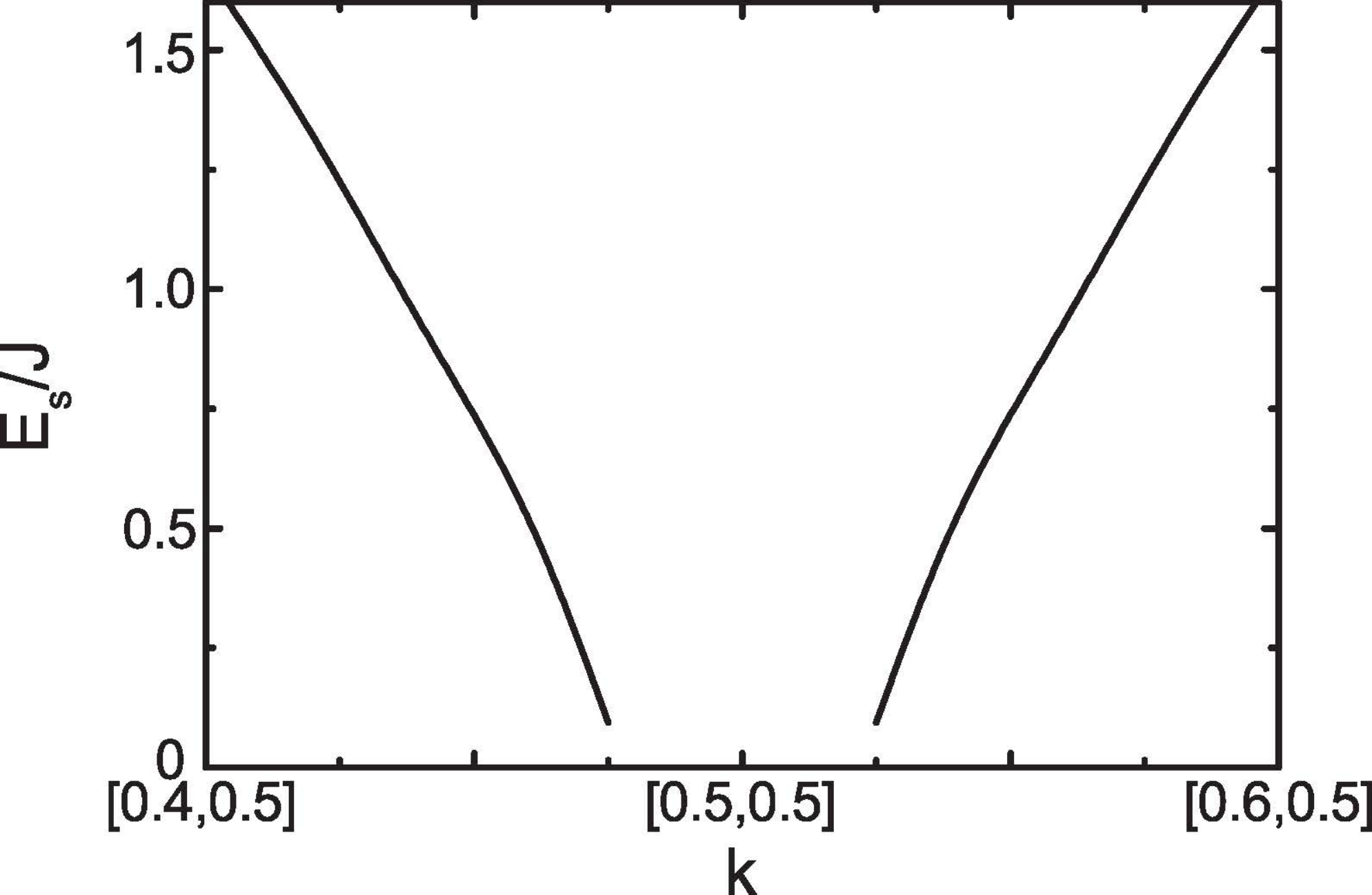}
\caption{The momentum dependence of the position of the low-energy magnetic scattering peaks in the normal-state at $\delta=0.15$ with $T=0.1J$ for $t/J=-2.5$ and $t'/t=0.3$. \label{NS-position}}
\end{figure}

\begin{figure*}[t!]
\centering
\includegraphics[scale=0.5]{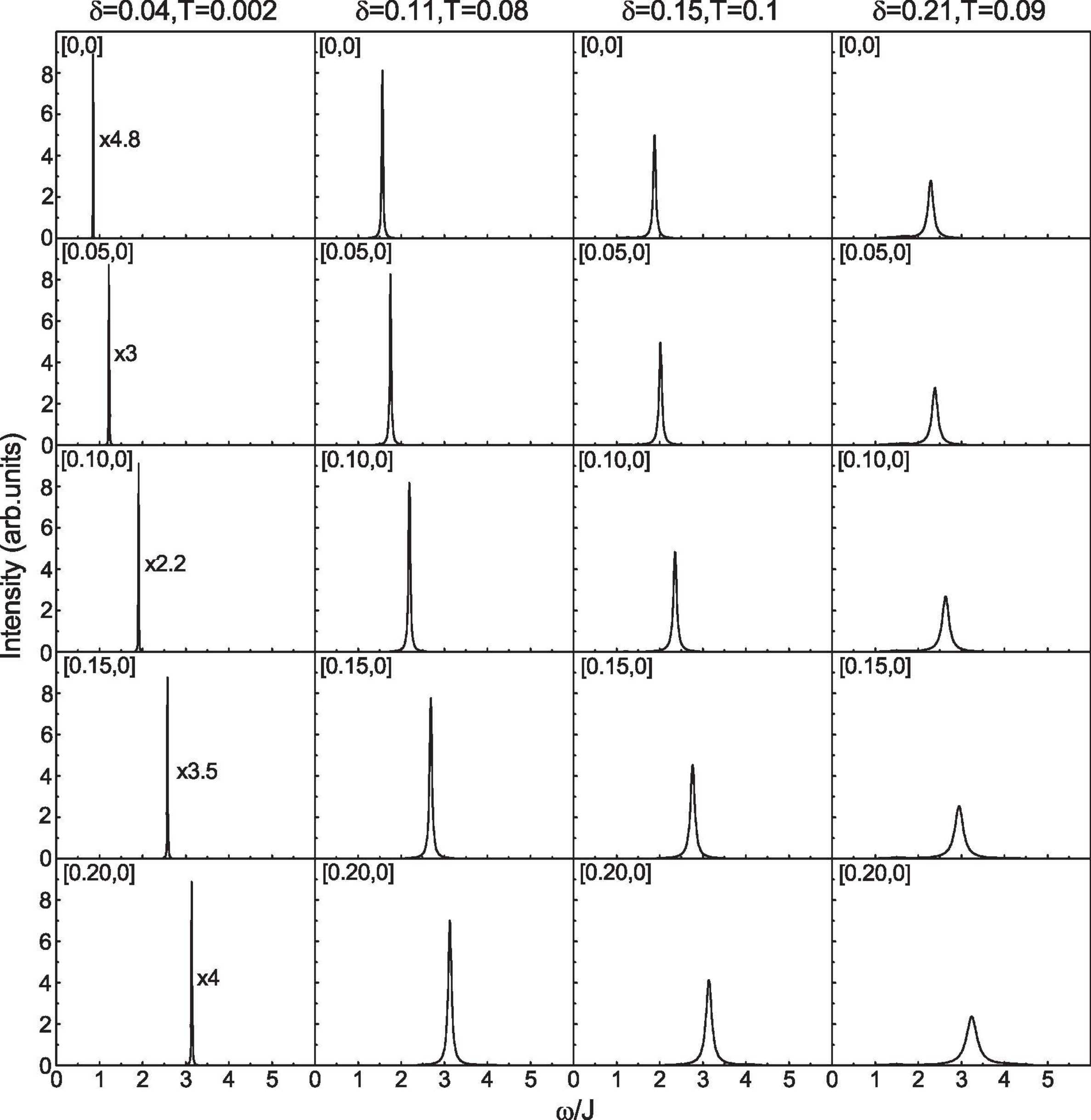}
\caption{The dynamical spin structure factor $S({\bf k},\omega)$ in the normal-state as a function of energy along the ${\bf k}=[0,0]$ to ${\bf k}=[0.5,0]$ direction of the Brillouin zone at $\delta=0.04$ with $T=0.002J$, $\delta=0.11$ with $T=0.08J$, $\delta=0.15$ $T=0.10J$, and $\delta=0.21$ with $T=0.09J$ for $t/J=-2.5$ and $t'/t=0.3$. \label{NS-high-energy}}
\end{figure*}

For the temperature above $T_{\rm c}$, i.e., $T>T_{\rm c}$, the system is in a normal-state, and then the SC-state $S({\bf k},\omega)$ in Eq. (\ref{DSSF}) is reduced to the case in the normal-state, where the spin self-energy is obtained in terms of the collective charge-carrier mode in the particle-hole channel {\it only}. To show how the spin excitations evolve with temperature from the SC-state to the normal-state, we firstly plot the positions of the low-energy magnetic scattering peaks around $[1/2,1/2]$ as a function of momentum at $\delta=0.15$ with $T=0.10J$ in Fig. \ref{NS-position}. Comparing it with Fig. \ref{position} for the same set of parameters except for the temperature $T=0.10J$, we see that the commensurate resonance is absent from the normal-state, reflecting a fact that the low-energy IC magnetic scattering in the SC-state can persist into the normal-state, while the low-energy commensurate resonance exists only in the SC-state. Since the height of the magnetic scattering peak is determined by damping, the IC magnetic scattering peak broadens and weakens in amplitude as the energy increase.

To analyze the evolution of the high-energy spin excitations with momentum and doping in the normal-state, we have performed a calculation of the normal-state $S({\bf k},\omega)$ at the different doping levels with the temperature well above the corresponding $T_{\rm c}$, and the result of the normal-state $S({\bf k},\omega)$ as a function of energy along the ${\bf k}=[0,0]$ to ${\bf k}= [0.5,0]$ direction of BZ at $\delta=0.04$ with $T=0.002J$, $\delta=0.11$ with $T=0.08J$, $\delta=0.15$ with $T=0.10J$, and $\delta=0.21$ with $T=0.09J$ is plotted in Fig. \ref{NS-high-energy}. In comparison with the corresponding result of the SC-state in Fig. \ref{SC-high-energy} for the same set of parameters except for the temperature above $T_{\rm c}$, we therefore find that these normal-state high-energy spin excitations, in their spectral weights and shapes of the magnetic scattering peaks, are striking similar to those in the corresponding SC-state, also in qualitative agreement with the experimental result \cite{Ishii14}. These results in Fig. \ref{SC-high-energy} and Fig. \ref{NS-high-energy} also indicate that the high-energy spin excitations do not correlate with the pairing interaction, and then the low-energy spin excitations are dominant mediating glue for the pairing in the framework of the kinetic-energy-driven superconductivity.

The essential physics of the electron-doped cuprate superconductors is the same as in the hole-doped case except for the electron-hole asymmetry. In the $t$-$J$ model, the NN hopping $t$ has a particle-hole symmetry because the sign of $t$ can be absorbed by the change of the sign of the orbital on one sublattice, however, the particle-hole asymmetry is described by the next NN hopping $t'$. Although there are the similar strengths of the AF exchange coupling $J$, NN hopping $t$, and next NN hopping $t'$ for both the hole- and electron-doped cuprate superconductors, the interplay of $t'$ with $t$ and $J$ causes a further reduction of the magnetic correlations in the hole-doped case, while increase of the AF spin fluctuations in the electron-doped side \cite{Hybertson90,Gooding94,Pavarini01}, i.e., the AF spin fluctuations in the electron-doped side are stronger than that in the hole-doped case, leading that the spin excitations in the electron-doped cuprate superconductors more deeply couple to the charge carriers. In this case, the hot spots are located much closer to the zone diagonal \cite{Matsui05}, and then the charge-carrier pair gap has a nonmonotonic d-wave gap form (\ref{gap}) as mentioned above. Although the momentum dependence of the SC-gap in Eq. (\ref{gap}) is basically consistent with the d-wave symmetry, it obviously deviates from the monotonic d-wave SC gap \cite{Matsui05}. A natural question is what is the reason why there are similarities and differences of the spin excitation spectrum between the electron- and hole-doped cuprate superconductors? To our present understanding, there are at least three reasons: (A) The dynamical spin response is manifested itself by the dynamical spin structure factor $S({\bf k},\omega)$ in Eq. (\ref{DSSF}), where the spin self-energy $\Sigma^{({\rm s})}({\bf k},\omega)$ in Eq. (\ref{SSF}) in the SC-state is obtained by including the contributions of the collective modes in both the particle-particle and particle-hole channels. However, as we \cite{Kuang15} have shown in the hole-doped case that the charge-carrier quasiparticle spectrum in the $t$-$J$ model has an effective band width $W\sim 2J$. This is also true in the electron-doped side, and then the spin self-energy in Eq. (\ref{SSF}) strongly renormalizes the spin excitations of the electron-doped cuprate superconductors at energies below $W$, but has a weak effect on the spin excitations at energies above $W$. This is why the dispersion of the spin excitations at energies below $W$ are strongly reorganized by electron doping, while the high-energy spin excitations are insensitive to electron doping; (B) On the other hand, the charge-carrier quasiparticle contribution to the spin self-energy renormalization in Eq. (\ref{SSF}) is separated as two parts $\Sigma^{({\rm s})}_{\rm ph}({\bf k},\omega)$ and $\Sigma^{({\rm s})}_{\rm pp}({\bf k},\omega)$, respectively. The first part of the contribution $\Sigma^{({\rm s})}_{\rm ph}({\bf k},\omega)$ in Eq. (\ref{SSF-ph}) is generated mainly by the mobile charge-carrier quasiparticles, and the coherence factor for this process is given in Eq. (\ref{CFS1}). In particular, it is straightforward to find when ${\bf k}=[1/2,1/2]$, $\Sigma^{({\rm s})}_{\rm ph}({\bf k},\omega)|_{{\bf k}=[1/2,1/2]}\approx 0$, reflecting a fact that this process mainly induces the low-energy IC magnetic scattering, and this low-energy IC magnetic scattering can persist into the normal-state. However, the second part of the contribution $\Sigma^{({\rm s})}_{\rm pp}({\bf k},\omega)$ in Eq. (\ref{SSF-pp}) originates from the creation of charge carrier pairs, and the coherence factor for this process is given in Eq. (\ref{CFS2}). When the charge-carrier pair gap parameter $\bar{\Delta}_{\rm a}=0$ for the temperature above $T_{\rm c}$, $\Sigma^{({\rm s})}_{\rm pp}({\bf k},\omega)|_{T>T_{\rm c}}=0$, indicating that this contribution occurs only in the SC-state, and therefore gives a dominant contribution to the commensurate magnetic scattering \cite{Cheng08}. As we \cite{Cheng08} have shown that the higher harmonic term in Eq. (\ref{gap}) mainly effects the low-energy behavior of the spin self-energy, i.e., the nonmonotonic d-wave SC gap in Eq. (\ref{gap}) in the electron-doped cuprate superconductors modulates the spin excitation spectrum in the electron-doped cuprate superconductors in terms of the spin self-energy, which induces a broad commensurate magnetic scattering at low energies in the SC-state \cite{Cheng08}, and therefore is different from the hole-doped case \cite{Kuang15}. This origin of the absence of the low-energy hour-glass-shaped dispersion is consistent with the discussions in Ref. \onlinecite{Ismer07}, where the differences in the low-energy spin resonances and their dispersions for the electron- and the hole-doped cuprate superconductors are attributed to the effect of the electron dispersion and the Fermi surface topology asymmetry between the electron- and hole-doped cuprate superconductors, and the higher harmonics in the d-wave gap on the electron-doped side. Moreover, we find that $\Sigma^{({\rm s})}_{\rm pp} ({\bf k},\omega)|_{{\bf k}=[1/2,1/2]}\approx 2\bar{\Delta}_{\rm a}$, this is why the magnetic resonance energy is intriguingly related to the SC gap \cite{Yu09}. In the kinetic-energy-driven SC mechanism \cite{Feng15,Feng0306}, superconductivity is mediated by the spin excitations, where the charge-carrier pair gap parameter $\bar{\Delta}_{\rm a}$ has dome-shaped doping dependence \cite{Cheng08}, which leads to that the resonance energy $\omega_{\rm r}$ shows the same dome-shaped doping dependence; (C) In contrast to the case in the SC-state, the spin self-energy in the normal-state is due to the charge carrier bubble in the charge-carrier particle-hole channel {\it only}, i.e., only process from the mobile charge-carrier quasiparticles contributes to the normal-state spin self-energy. This difference leads to an absence of the commensurate resonance in the normal-state, and therefore further confirm that the commensurate resonance appears in the SC-state only, while the low-energy IC magnetic scattering can persist into the normal-state. Moreover, the effective band width $W$ of the charge-carrier quasiparticle spectrum in the normal-state is the same as that in the corresponding SC-state, and then in analogy to the case in the SC-state, the spin self-energy strongly renormalizes the spin excitations at energies below $W$, but has a weak effect on the spin excitations at energies above $W$. This is why the high-energy spin excitations in the normal-state retain roughly constant energy as a function of doping, with the shape of the magnetic scattering peaks, spectral weights and dispersion relations comparable to those in the corresponding SC-state.

\section{Summary and discussion}\label{conclusions}

In summary, within the framework of the kinetic-energy-driven SC mechanism, we have studied the momentum and doping dependence of the dynamical spin response in the electron-doped cuprate superconductors. Our results show that the dispersion of the low-energy spin excitations depend sensitively on doping and momentum. However, the hour-glass-shaped dispersion of the low-energy spin excitations appeared in the hole-doped case is absent in the electron-doped side due to the electron-hole asymmetry. In particular, although the low-energy IC magnetic scattering can persist into the normal-state, the broad low-energy commensurate magnetic scattering appears only in the SC-state, with the commensurate resonance energy that correlates with the dome-shaped doping dependence of the charge-carrier pair gap parameter. Moreover, the spectral weight and dispersion of the high-energy spin excitations in the SC-state are comparable with those in the corresponding normal-state. Incorporating the present result with that obtained in the hole-doped cuprate superconductors \cite{Kuang15}, it is thus shown that the high-energy spin excitations do not correlate with the pairing interaction, while the spectral weight and dispersion at low-energy in establishing a relevant energy scale and strength of the spin excitations is important for pairing.

\acknowledgments

PJ, LK, and SF are supported by the National Key Research and Development Program of China under Grant No. 2016YFA0300304, and National Natural Science Foundation of China (NSFC) under Grant Nos. 11274044 and 11574032, HZ is supported by NSFC under Grant No. 11547034, and YL is supported by the Science Foundation of Hengyang Normal University under Grant No. 13B44, and Hunan Provincial Natural Science Foundation of China under Grant No. 2015JJ3027.

\end{document}